\pdfoutput=1
\documentclass[conference, a4paper]{IEEEtran}

\usepackage[T1]{fontenc}
\usepackage[utf8]{inputenc}
\usepackage[english]{babel}
\usepackage{amsmath, amsthm, amssymb}

\usepackage{hyperref}

\usepackage{multirow}
\usepackage{csquotes}
\usepackage{nicefrac}
\usepackage[textsize=tiny]{todonotes}
\usepackage{booktabs}

\usepackage{tikz}
\usetikzlibrary{shapes.geometric, positioning, calc, matrix}

\usepackage{physics}
\usetikzlibrary{quantikz}

\usepackage{subcaption}

\usepackage[style=ieee, maxnames=3, minnames=1, date=year, doi=false,isbn=false,backend=biber]{biblatex}
\AtEveryBibitem{
  \clearname{editor}%
  \clearfield{series}%
  \clearfield{isbn}%
  \clearfield{issn}%
  \clearfield{volume}
  \clearfield{number}
  \clearfield{pages}
}

\usepackage[binary-units=true, detect-all=true]{siunitx}
\usepackage{etoolbox}
\robustify\bfseries

\usepackage{nicematrix}
\usepackage{flushend}

\newtheorem{example}{Example}

\definecolor{RedEdge}{RGB}{191,40,40}
\definecolor{BlueEdge}{RGB}{40,191,191}
\tikzset{%
	terminal/.style={draw,rectangle,inner sep=2pt,font=\footnotesize,very thick},
	zeronode/.style={fill, draw, circle, minimum width=2pt, inner sep=0pt,color=black},
	qubit/.style={draw,circle,inner sep=0pt,minimum width=0.35cm,minimum height=0.35cm,font=\footnotesize, thin},
	edgeOne/.style={color=RedEdge,ultra thick},
	edgeOneState/.style={color=RedEdge, thick},
	edgeMOne/.style={color=BlueEdge,ultra thick},
	edgeSqrt/.style={color=RedEdge, thick},
	edgeMSqrt/.style={color=BlueEdge, thick},
	edgeFrac/.style={color=RedEdge, thin},
	edgeOver/.style={dotted, color=blue, ultra thick}
}

\tikzset{%
	font={\footnotesize},
	terminal/.style={draw,rectangle,inner sep=2pt,font=\footnotesize,very thick},
	vertex/.style={draw,circle,inner sep=0pt,minimum width=0.5cm,minimum height=0.5cm},
	define color/.code={\definecolor{hsb#1}{Hsb}{#1, 1, 0.75}},				
    medge/.style n args={3}{
		line width={#1pt},
		define color={#2},
		draw=hsb#2,
		out=#3, 
		in=90
	},
    edge/.style 2 args={
		line width={#1pt},
		define color={#2},
		draw=hsb#2
	},
	edge0/.style 2 args={
		line width={#1pt},
		define color={#2},
		draw=hsb#2,
		out=-130, 
		in=90
	},
	edge1/.style 2 args={
		line width={#1pt},
		define color={#2},
		draw=hsb#2,
		out=-50, 
		in=90
	},
	zerostub/.style={
		inner sep=0, 
		minimum size=3pt, 
		circle, 
		fill=black
	}
}

\newsavebox{\CXgateb}
\savebox{\CXgateb}{%
	\begin{tikzpicture}
		\matrix[ampersand replacement=\&,column sep={0.3cm,between origins},row sep={0.6cm,between origins}] (qmdd) {
			\& \node[qubit] (q2) {$2$}; \& \\
			\& \node[qubit] (q1) {$1$};         \& \\
			\node[qubit] (q0a) {$0$};  \&                             \& \node[qubit] (q0b) {$0$}; \\
			\& \node[terminal] (t) {}; \& \\
		};
		
		\draw[edgeOne] ($(q2)+(0,0.5cm)$) -- (q2);
		
		\draw[edgeOne] (q2) to[out=-150,in=150]   (q1);
		\draw[edgeOne] (q2) to[out=-110,in=90] ++(250:0.3cm)  node[zeronode] {};			
		\draw[edgeOne] (q2) to[out=-70,in=90]  ++(290:0.3cm)  node[zeronode] {};
		\draw[edgeOne] (q2) to[out=-30,in=30]    (q1);
		
		\draw[edgeOne] (q1) to[out=-150,in=90]   (q0a);
		\draw[edgeOne] (q1) to[out=-110,in=90] ++(250:0.3cm)  node[zeronode] {};			
		\draw[edgeOne] (q1) to[out=-70,in=90]  ++(290:0.3cm)  node[zeronode] {};			
		\draw[edgeOne] (q1) to[out=-30,in=90]    (q0b);
		
		\draw[edgeOne] (q0a) to[out=-150,in=180] (t);
		\draw[edgeOne] (q0a) to[out=-110,in=90]  ++(250:0.3cm) node[zeronode] {};			
		\draw[edgeOne] (q0a) to[out=-70,in=90]   ++(290:0.3cm)  node[zeronode] {};
		\draw[edgeOne] (q0a) to[out=-30,in=90]   (t);

		\draw[edgeOne] (q0b) to[out=-150,in=90]  ++(220:0.3cm) node[zeronode] {};
		\draw[edgeOne] (q0b) to[out=-110,in=60] (t);
		\draw[edgeOne] (q0b) to[out=-70,in=0]   (t);		
		\draw[edgeOne] (q0b) to[out=-30,in=90]   ++(320:0.3cm)  node[zeronode] {};
	\end{tikzpicture}
}

\newsavebox{\CXgatemqubits}
\savebox{\CXgatemqubits}{%
	\begin{tikzpicture}[every node/.style={scale=0.8}]
		\matrix[ampersand replacement=\&,column sep={0.6cm,between origins},row sep={0.9cm,between origins}] (qmdd) {
			\& \node[qubit] (q2) {$n-1$}; \& \\
			\node[qubit] (q1a) {$n-2$};  \&                             \& \node[qubit] (q1b) {$n-2$}; \\
			\node[] (z1a) {$\vdots$}; \& \& \node[] (z1b) {$\vdots$}; \\
			\node[qubit, minimum size = 0.6cm] (q0a) {$0$};  \&                             \& \node[qubit, minimum size = 0.6cm] (q0b) {$0$}; \\
			\& \node[terminal] (t) {}; \& \\
		};
		
		\draw[edgeOne] ($(q2)+(0,0.5cm)$) -- (q2);
		
		\draw[edgeOne] (q2) to[out=-150,in=90]   (q1a);
		\draw[edgeOne] (q2) to[out=-110,in=90] ++(250:0.6cm)  node[zeronode] {};			
		\draw[edgeOne] (q2) to[out=-70,in=90]  ++(290:0.6cm)  node[zeronode] {};
		\draw[edgeOne] (q2) to[out=-30,in=90]    (q1b);
		
		\draw[edgeOne] (q1a) to[out=-150,in=150]   ++(-0.3,-0.6);
		\draw[edgeOne] (q1a) to[out=-110,in=90] ++(250:0.5cm)  node[zeronode] {};			
		\draw[edgeOne] (q1a) to[out=-70,in=90]  ++(290:0.5cm)  node[zeronode] {};			
		\draw[edgeOne] (q1a) to[out=-30,in=30]    ++(0.3,-0.6);
		
		\draw[edgeOne] (q1b) to[out=-150,in=150]   ++(-0.3,-0.6);
		\draw[edgeOne] (q1b) to[out=-110,in=90] ++(250:0.5cm)  node[zeronode] {};			
		\draw[edgeOne] (q1b) to[out=-70,in=90]  ++(290:0.5cm)  node[zeronode] {};			
		\draw[edgeOne] (q1b) to[out=-30,in=30]    ++(0.3,-0.6);
		
		\draw[edgeOne] (q0a) to[out=130,in=-150]   ++(-0.3,0.6);
		\draw[edgeOne] (q0a) to[out=50,in=-30]   ++(0.3,0.6);		
		\draw[edgeOne] (q0a) to[out=-150,in=180] (t);
		\draw[edgeOne] (q0a) to[out=-110,in=90]  ++(250:0.5cm) node[zeronode] {};			
		\draw[edgeOne] (q0a) to[out=-70,in=90]   ++(290:0.5cm)  node[zeronode] {};
		\draw[edgeOne] (q0a) to[out=-30,in=90]   (t);
		
		\draw[edgeOne] (q0b) to[out=130,in=-150]   ++(-0.3,0.6);
		\draw[edgeOne] (q0b) to[out=50,in=-30]   ++(0.3,0.6);
		\draw[edgeOne] (q0b) to[out=-150,in=90]  ++(220:0.5cm) node[zeronode] {};
		\draw[edgeOne] (q0b) to[out=-110,in=60] (t);
		\draw[edgeOne] (q0b) to[out=-70,in=0]   (t);		
		\draw[edgeOne] (q0b) to[out=-30,in=90]   ++(320:0.5cm)  node[zeronode] {};

	\end{tikzpicture}
}

\newsavebox{\CXgatemulti}
\savebox{\CXgatemulti}{%
	\begin{tikzpicture}[every node/.style={scale=0.8}]
		\matrix[ampersand replacement=\&,column sep={0.6cm,between origins},row sep={0.9cm,between origins}] (qmdd) {
			\& \node[qubit] (q3) {$n-1$}; \& \\
			\node[qubit] (q2a) {$n-2$};  \&                             \& \node[qubit] (q2b) {$n-2$}; \\
			\node[qubit] (q1a) {$n-3$};  \&                             \& \node[qubit] (q1b) {$n-3$}; \\
			\node[] (z1a) {$\vdots$}; \& \& \node[] (z1b) {$\vdots$}; \\
			\node[qubit, minimum size = 0.6cm] (q0a) {$0$};  \&                             \& \node[qubit, minimum size = 0.6cm] (q0b) {$0$}; \\
			\& \node[terminal] (t) {}; \& \\
		};
		
		\draw[edgeOne] ($(q3)+(0,0.5cm)$) -- (q3);
		
		\draw[edgeOne] (q3) to[out=-150,in=90]   (q2a);
		\draw[edgeOne] (q3) to[out=-110,in=90] ++(250:0.6cm)  node[zeronode] {};			
		\draw[edgeOne] (q3) to[out=-70,in=90]  ++(290:0.6cm)  node[zeronode] {};
		\draw[edgeOne] (q3) to[out=-30,in=90]    (q2b);		
		
		\draw[edgeOne] (q2a) to[out=-150,in=150]   (q1a);
		\draw[edgeOne] (q2a) to[out=-110,in=90] ++(250:0.5cm)  node[zeronode] {};			
		\draw[edgeOne] (q2a) to[out=-70,in=90]  ++(290:0.5cm)  node[zeronode] {};
		\draw[edgeOne] (q2a) to[out=-30,in=50]    (q1a);
		
		\draw[edgeOne] (q2b) to[out=-150,in=30]   (q1a);
		\draw[edgeOne] (q2b) to[out=-110,in=90] ++(250:0.5cm)  node[zeronode] {};			
		\draw[edgeOne] (q2b) to[out=-70,in=90]  ++(290:0.5cm)  node[zeronode] {};
		\draw[edgeOne] (q2b) to[out=-30,in=50]    (q1b);
		
		\draw[edgeOne] (q1a) to[out=-150,in=150]   ++(-0.3,-0.6);
		\draw[edgeOne] (q1a) to[out=-110,in=90] ++(250:0.5cm)  node[zeronode] {};			
		\draw[edgeOne] (q1a) to[out=-70,in=90]  ++(290:0.5cm)  node[zeronode] {};			
		\draw[edgeOne] (q1a) to[out=-30,in=30]    ++(0.3,-0.6);
		
		\draw[edgeOne] (q1b) to[out=-150,in=30]   ++(-0.9,-0.6);
		\draw[edgeOne] (q1b) to[out=-110,in=90] ++(250:0.5cm)  node[zeronode] {};			
		\draw[edgeOne] (q1b) to[out=-70,in=90]  ++(290:0.5cm)  node[zeronode] {};			
		\draw[edgeOne] (q1b) to[out=-30,in=30]    ++(0.3,-0.6);
		
		\draw[edgeOne] (q0a) to[out=130,in=-150]   ++(-0.3,0.6);
		\draw[edgeOne] (q0a) to[out=50,in=-30]   ++(0.3,0.6);
		\draw[edgeOne] (q0a) to[out=40,in=-150]   ++(0.9,0.6);
		\draw[edgeOne] (q0a) to[out=-150,in=180] (t);
		\draw[edgeOne] (q0a) to[out=-110,in=90]  ++(250:0.5cm) node[zeronode] {};			
		\draw[edgeOne] (q0a) to[out=-70,in=90]   ++(290:0.5cm)  node[zeronode] {};
		\draw[edgeOne] (q0a) to[out=-30,in=90]   (t);
		
		\draw[edgeOne] (q0b) to[out=50,in=-30]   ++(0.3,0.6);
		\draw[edgeOne] (q0b) to[out=-150,in=90]  ++(220:0.5cm) node[zeronode] {};
		\draw[edgeOne] (q0b) to[out=-110,in=60] (t);
		\draw[edgeOne] (q0b) to[out=-70,in=0]   (t);		
		\draw[edgeOne] (q0b) to[out=-30,in=90]   ++(320:0.5cm)  node[zeronode] {};

	\end{tikzpicture}
}

\newsavebox{\CXgatea}
\savebox{\CXgatea}{%
	\begin{tikzpicture}
		\matrix[ampersand replacement=\&,column sep={0.3cm,between origins},row sep={0.6cm,between origins}] (qmdd) {
			\& \node[qubit] (q2) {$2$}; \& \\
			\node[qubit] (q1a) {$1$};  \&                             \& \node[qubit] (q1b) {$1$}; \\
			\& \node[qubit] (q0) {$0$};         \& \\
			\& \node[terminal] (t) {}; \& \\
		};
		
		\draw[edgeOne] ($(q2)+(0,0.5cm)$) -- (q2);
		
		\draw[edgeOne] (q2) to[out=-150,in=90]   (q1a);
		\draw[edgeOne] (q2) to[out=-110,in=90] ++(250:0.3cm)  node[zeronode] {};			
		\draw[edgeOne] (q2) to[out=-70,in=90]  ++(290:0.3cm)  node[zeronode] {};
		\draw[edgeOne] (q2) to[out=-30,in=90]    (q1b);
		
		\draw[edgeOne] (q1a) to[out=-150,in=-180]   (q0);
		\draw[edgeOne] (q1a) to[out=-110,in=90] ++(250:0.3cm)  node[zeronode] {};			
		\draw[edgeOne] (q1a) to[out=-70,in=90]  ++(290:0.3cm)  node[zeronode] {};			
		\draw[edgeOne] (q1a) to[out=-30,in=90]    (q0);
		
		\draw[edgeOne] (q1b) to[out=-150,in=90]  ++(220:0.3cm) node[zeronode] {};
		\draw[edgeOne] (q1b) to[out=-110,in=60] (q0);
		\draw[edgeOne] (q1b) to[out=-70,in=0]   (q0);		
		\draw[edgeOne] (q1b) to[out=-30,in=90]   ++(320:0.3cm)  node[zeronode] {};

		\draw[edgeOne] (q0) to[out=-150,in=-180] (t);
		\draw[edgeOne] (q0) to[out=-110,in=90] ++(250:0.3cm)  node[zeronode] {};
		\draw[edgeOne] (q0) to[out=-70,in=90]  ++(290:0.3cm)  node[zeronode] {};			
		\draw[edgeOne] (q0) to[out=-30,in=0]   (t);
	\end{tikzpicture}
}

\newsavebox{\Hgate}
\sbox{\Hgate}{%
	\begin{tikzpicture}
		\matrix[ampersand replacement=\&,column sep={0.3cm,between origins},row sep={0.6cm,between origins}] (qmdd) {
			\& \node[qubit] (q2) {$2$}; \& \\
			\& \node[qubit] (q1) {$1$};  \&   \\
			\& \node[qubit] (q0) {$1$};         \& \\
			\& \node[terminal] (t) {}; \& \\
		};
		\draw[edgeFrac] ($(q2)+(0,0.5cm)$) -- (q2);
		
		\draw[edgeOne] (q2) to[out=-150,in=150]   (q1);
		\draw[edgeOne] (q2) to[out=-120,in=120] (q1);			
		\draw[edgeOne] (q2) to[out=-60,in=60]  (q1);
		\draw[edgeMOne] (q2) to[out=-30,in=30]    (q1);
		
		\draw[edgeOne] (q1) to[out=-150,in=-180]   (q0);
		\draw[edgeOne] (q1) to[out=-110,in=90] ++(250:0.3cm)  node[zeronode] {};			
		\draw[edgeOne] (q1) to[out=-70,in=90]  ++(290:0.3cm)  node[zeronode] {};			
		\draw[edgeOne] (q1) to[out=-30,in=0]    (q0);

		\draw[edgeOne] (q0) to[out=-150,in=-180] (t);
		\draw[edgeOne] (q0) to[out=-110,in=90] ++(250:0.3cm)  node[zeronode] {};
		\draw[edgeOne] (q0) to[out=-70,in=90]  ++(290:0.3cm)  node[zeronode] {};			
		\draw[edgeOne] (q0) to[out=-30,in=0]   (t);
	\end{tikzpicture}
}

\newsavebox{\ddfullstate}
\sbox{\ddfullstate}{%
	\begin{tikzpicture}
		\matrix[ampersand replacement=\&,column sep={0.4cm,between origins},row sep={0.7cm,between origins}] (qmdd) {
			\& \node[qubit] (q2) {$2$}; \& \\
			\node[qubit] (q1a) {$1$};  \&                             \& \node[qubit] (q1b) {$1$}; \\
			\node[qubit] (q0a) {$1$};  \&                             \& \node[qubit] (q0b) {$1$}; \\
			\& \node[terminal] (t) {}; \& \\
		};
		
		\draw[edgeOne] ($(q2)+(0,0.55cm)$) -- (q2);
		\draw[edgeOver] ($(q2)+(0,0.55cm)$) -- (q2);
		
		\draw[edgeFrac] (q2) to[out=-150,in=90] (q1a);
		\draw[edgeOver] (q2) to[out=-150,in=90] (q1a);			
		\draw[edgeFrac] (q2) to[out=-30,in=90] (q1b);
		
		\draw[edgeOne] (q1a) to[out=-150,in=140] (q0a);
		\draw[edgeOver] (q1a) to[out=-150,in=140] (q0a);		
		\draw[edgeOne] (q1a) to[out=-30,in=100] ++(320:0.3cm)  node[zeronode] {};
		
		\draw[edgeOne] (q1b) to[out=-150,in=80] ++(220:0.3cm)  node[zeronode] {};			
		\draw[edgeOne] (q1b) to[out=-30,in=40] (q0b);
		
		\draw[edgeOne] (q0a) to[out=-150,in=140] (t);
		\draw[edgeOver] (q0a) to[out=-150,in=140] (t);				
		\draw[edgeOne] (q0a) to[out=-30,in=100]  ++(320:0.3cm)  node[zeronode] {};
		
		\draw[edgeOne] (q0b) to[out=-150,in=80] ++(220:0.3cm)  node[zeronode] {};			
		\draw[edgeOne] (q0b) to[out=-30,in=40]  (t);
	\end{tikzpicture}
}

\newsavebox{\ddinitialstate}
\sbox{\ddinitialstate}{%
	\begin{tikzpicture}
		\matrix[ampersand replacement=\&,column sep={0.4cm,between origins},row sep={0.7cm,between origins}] (qmdd) {
			\& \node[qubit] (q2) {$2$}; \& \\
			\& \node[qubit] (q1) {$1$}; \& \\
			\& \node[qubit] (q0) {$0$}; \& \\
			\& \node[terminal] (t) {}; \& \\
		};
		
		\draw[edgeOne] ($(q2)+(0,0.55cm)$) -- (q2);
		
		\draw[edgeOne] (q2) to[out=-150,in=140] (q1);			
		\draw[edgeOne] (q2) to[out=-30,in=100] ++(320:0.3cm)  node[zeronode] {};
		
		\draw[edgeOne] (q1) to[out=-150,in=140] (q0);			
		\draw[edgeOne] (q1) to[out=-30,in=100] ++(320:0.3cm)  node[zeronode] {};
		
		\draw[edgeOne] (q0) to[out=-150,in=140] (t);			
		\draw[edgeOne] (q0) to[out=-30,in=100]  ++(320:0.3cm)  node[zeronode] {};
	\end{tikzpicture}
}

\begin{document}

\title{Tensor Networks or Decision Diagrams?\\ {\LARGE Guidelines for Classical Quantum Circuit Simulation}}

\author{
	\IEEEauthorblockN{Lukas Burgholzer\IEEEauthorrefmark{1}\hspace*{1.5cm}Alexander Ploier\IEEEauthorrefmark{1}\hspace*{1.5cm}Robert Wille\IEEEauthorrefmark{2}\IEEEauthorrefmark{3}}
	\IEEEauthorblockA{\IEEEauthorrefmark{1}Institute for Integrated Circuits, Johannes Kepler University Linz, Austria}
	\IEEEauthorblockA{\IEEEauthorrefmark{2}Chair for Design Automation, Technical University of Munich, Germany}
	\IEEEauthorblockA{\IEEEauthorrefmark{3}Software Competence Center Hagenberg GmbH (SCCH), Austria}
	\IEEEauthorblockA{\href{mailto:lukas.burgholzer@jku.at}{lukas.burgholzer@jku.at}\hspace{1.5cm}\href{mailto:alexander.ploier@jku.at}{alexander.ploier@jku.at}\hspace{1.5cm} \href{mailto:robert.wille@tum.de}{robert.wille@tum.de}\\
	\url{https://www.cda.cit.tum.de/research/quantum/}}
	\vspace*{-0.5cm}
}

\maketitle

\begin{abstract}
Classically simulating quantum circuits is crucial when developing or testing quantum algorithms. 
Due to the underlying exponential complexity, efficient data structures are key for performing such simulations.
To this end, \emph{tensor networks} and \emph{decision diagrams} have independently been developed with differing perspectives, terminologies, and backgrounds in mind. 
Although this left designers with two complementary data structures for quantum circuit simulation, thus far it remains unclear which one is the better choice for a given use case.
In this work, we $(1)$ consider how these techniques approach classical quantum circuit simulation, and $(2)$ examine their (dis)similarities with regard to their most applicable \emph{abstraction level}, the \emph{desired simulation output}, the impact of the \emph{computation order}, and the ease of \emph{distributing the workload}. 
As a result, we provide guidelines for when to better use tensor networks and when to better use decision diagrams in classical quantum circuit simulation.
\end{abstract}

\section{Introduction}
\label{sec:introduction}
The current state of quantum computing often prohibits evaluating prototypes or applications on the actual hardware, due to its cost, reliability, and availability. 
Therefore, classically simulating quantum circuits is an integral part in the development and testing process of quantum algorithms/applications---providing
meaningful insights into the functionality and performance of a quantum system by allowing to access the complete, otherwise inaccessible, quantum state and (deterministically) studying the effects of errors/noise.

Performing a quantum computation entails evolving an initial quantum state by applying a sequence of operations (also called gates) that is described as a quantum circuit.
The goal of classically simulating such a computation is to calculate a representation of the evolved system state (or at least individual amplitudes thereof).
To this end, the state of a quantum system can be described as a vector of complex amplitudes, while individual operations can be represented by unitary matrices.
Thus, classically simulating a quantum circuit essentially translates to multiplications of vectors and matrices.  

However, while conceptually simple, the respective vectors and matrices grow exponentially with respect to the number of qubits.
For example, the full state of a $32$-qubit system is described by $2^{32}=\num{4294967296}$ complex amplitudes and storing it would require \SI{64}{\gibi\byte} of memory (assuming a \SI{128}{\bit} representation for complex numbers).
Consequently, quantum circuit simulation on classical machines in a straight-forward fashion quickly becomes prohibitive for larger systems---even when resorting to powerful supercomputing clusters.

Efficient data structures are necessary to tackle the inherent complexity of quantum mechanical systems.
Due to quantum computing being a multidisciplinary field involving physicists, computer scientists, and many others, multiple perspectives to this challenge exist with differing motivations, terminologies, and backgrounds in mind.
Inspired by the way nature behaves at the scale of atoms and sub-atomic particles, \emph{tensor networks}~\cite{fannesFinitelyCorrelatedStates1992, biamonteTensorNetworksNutshell2017, bridgemanHandwavingInterpretiveDance2017} naturally arose as a means to efficiently represent quantum mechanical systems back in~1992.
Much more recently, inspired by their success in the design of classical circuits and systems, \emph{decision diagrams}~\cite{niemannQMDDsEfficientQuantum2016, chin-yungExtendedXQDDRepresentation2011, zulehnerHowEfficientlyHandle2019, viamontesHighperformanceQuIDDBasedSimulation2004} have independently been proposed as a means to compactly represent and efficiently manipulate quantum functionality by exploiting redundancies in the underlying representations.

In this work, we $(1)$ consider how these techniques, that originated from different backgrounds, tackle the complexity inherent to the classical simulation of quantum circuits, and $(2)$ systematically analyze their (dis)similarities with regard to the following aspects:
\begin{itemize}
\item \textbf{Abstraction Level:} Several use cases for classically simulating quantum circuits on different abstraction layers exist, e.g., developing/testing of high-level applications or \mbox{low-level} verification of real quantum computers. The question is, where tensor networks and decision diagrams work best.
\item \textbf{Desired Simulation Output:} The desired output of a classical simulation can be anything from a scalar quantity to the complete state vector. It is key to understand what the limitations of the respective techniques are.
\item \textbf{Computation Order:} 
The performance of either technique heavily depends on the order in which the individual operations are conducted. 
While there is an immediate duality between the task for tensor networks and decision diagrams, the question is, whether the respective techniques are interchangeable between domains.
\item \textbf{Distributing the Workload:} 
In order to perform classical simulations of large quantum circuits, the respective methods have to be scalable to supercomputers. 
This raises the question, how well both techniques allow to make use of the available resources.
\end{itemize}
Overall, these considerations lead to guidelines for choosing an appropriate data structure to classically simulate quantum circuits depending on the respective use cases.

The rest of this paper is structured as follows: \autoref{sec:quantum_circuit_simulation} provides the necessary background on classical quantum circuit simulation. Based on that,~\autoref{sec:two_perspectives_on_quantum_circuit_simulation} introduces and examines tensor networks and decision diagrams as two complementary approaches on how to tackle its inherent complexity.
Then, \autoref{sec:two_sides_of_the_same_coin?} compares both techniques according to the aspects introduced above.
Finally, \autoref{sec:conclusion} concludes the paper.

\section{Classical Quantum Circuit Simulation}
\label{sec:quantum_circuit_simulation}
To keep this paper self contained, this section gives a short review of the main concepts of classical quantum circuit simulation. While the descriptions are kept brief, we refer the interested reader to \cite{nielsenQuantumComputationQuantum2010} for further details.

In contrast to classical computing, the state $\ket{\varphi}$ of a qubit can be described as a complex, linear combination (or \emph{superposition}) of the basis states $\ket{0}$ and $\ket{1}$, i.e., 
\[
\ket{\varphi}=\alpha_0 \ket{0} + \alpha_1 \ket{1} \mbox{ with } \alpha_0, \alpha_1\in\mathbb{C} \mbox{ and } |\alpha_0|^2+|\alpha_1|^2=1.
\]
In general, the state~$\ket{\varphi}$ of an $n$-qubit system is described by $2^n$ complex amplitudes, i.e., 
\[
\ket{\varphi} = \alpha_{0\dots 0} \ket{0\dots 0} + \cdots + \alpha_{1\dots 1} \ket{1\dots 1} 
\]
with \mbox{$\alpha_{0\dots 0},\dots, \alpha_{1\dots 1}\in\mathbb{C}$} and \mbox{$|\alpha_{0\dots 0}|^2+\dots +|\alpha_{1\dots 1}|^2=1$}.
This is frequently represented as the column vector $[\alpha_{0\dots 0}, \dots, \alpha_{1\dots 1}]^\top$ and referred to as the \emph{state vector}.

This state can be manipulated by applying \textit{quantum operations} (also called \textit{quantum gates}).
To this end, an operation acting on $k$ qubits can be represented by a unitary matrix of size $2^k\times 2^k$.
Applying an operation then corresponds to computing the matrix-vector product of the operation's unitary matrix with the current state vector\footnote{Technically, for the multiplication to make sense, any operation acting only on $k<n$ qubits has to be extended to the full system size by forming appropriate tensor products with identity matrices before performing the multiplication.}.

A \emph{quantum circuit} $G$ is described as a sequence of quantum gates, i.e., $G=g_0,\dots,g_{|G|-1}$ with each gate $g_i$ being represented by a corresponding unitary matrix $U_i$.
Given an initial state $\ket{\varphi_{\mathit{init}}}$, performing the computation described by the circuit $G$ entails applying all operations of the circuit to the initial state, i.e., computing
\[
	\ket{\varphi_{\mathit{final}}} = U_{|G|-1}\times\cdots\times U_0 \ket{\varphi_{\mathit{init}}} = U \ket{\varphi_{\mathit{init}}}.
\]
If this task is conducted on a classical computer, it is commonly referred to as \emph{classical quantum circuit simulation}.
Due to the exponential nature of the underlying representations---states being described by $2^n$ amplitudes---one might only be interested in determining particular amplitudes of the final state. This entails the calculation of $\alpha_i = \matrixel{i}{U_{|G|-1}\times\cdots\times U_0}{\varphi_{\mathit{init}}} = \matrixel{i}{U}{\varphi_{\mathit{init}}}$, with $\bra{i}$ denoting a row vector representing the $i$-th basis state.

\begin{figure}[t]
	\centering
	\begin{subfigure}[b]{0.4\linewidth}
	\centering
	\resizebox{0.88\linewidth}{!}{
	\begin{tikzpicture}
	\node[scale=1.1] (G) {
	\begin{quantikz}[column sep=10pt, row sep={17.5pt,between origins}]
		\lstick{$q_2\colon\ket{0}$} & \gate[style={fill=blue!20}]{H}  & \ctrl{1} & \qw & \qw  \rstick{$\bra{0}$}\\
		\lstick{$q_1\colon\ket{0}$} & \qw  & \targ{} & \ctrl{1}      & \qw  \rstick{$\bra{0}$} \\
		\lstick{$q_0\colon\ket{0}$} & \qw & \qw      & \targ{} & \qw  \rstick{$\bra{0}$}
	\end{quantikz}
	};
	\end{tikzpicture}}
	\caption{Quantum circuit}
	\label{fig:ghz_circuit}
	\end{subfigure}%
	\hfill
	\begin{subfigure}[b]{0.59\linewidth}
	\centering
	\resizebox{0.99\linewidth}{!}{
	\begin{tikzpicture}
	\node[below = 2cm of G.165, minimum height = 0.15cm] (CXmatA) {
	$\begin{bNiceMatrix}[small]1 & & & & & & & \\ & 1 & & & & & & \\ & & 0 & 1 & & & & \\ & & 1 & 0 & & & & \\ & & & & 1 & & & \\ & & & & & 1 & & \\ & & & & & & 0 & 1 \\ & & & & & & 1 & 0 \end{bNiceMatrix}$
	};
	\node[right = -0.1cm of CXmatA, minimum height = 0.15cm] (CXmatB){
	$\begin{bNiceMatrix}[small]1 & & & & & & & \\ & 1 & & & & & & \\ & & 1 & & & & & \\ & & & 1 & & & & \\ & & & & 0 & 0 & 1 & 0 \\ & & & & 0 & 0 & 0 & 1 \\ & & & & 1 & 0 & 0 & 0 \\ & & & & 0 & 1 & 0 & 0 \end{bNiceMatrix}$
	};
	\node[right = -0.0cm of CXmatB, minimum height = 0.15cm] (Hmat){
	$\frac{1}{\sqrt{2}}\begin{bNiceMatrix}[small]1 & & & & 1 & & & \\ & 1 & & & & 1 & & \\ & & 1 & & & & 1 & \\ & & & 1 & & & & 1 \\ 1 & & & & -1 & & & \\ & 1 & & & & -1 & & \\ & & 1 & & & & -1 & \\ & & & 1 & & & & -1 \end{bNiceMatrix}$
	};
	\node[right = -0.1cm of Hmat, minimum height = 0.15cm] (Input){
	$\begin{bNiceMatrix}[small]1 \\ 0 \\ 0 \\ 0 \\ 0 \\ 0 \\ 0 \\ 0 \end{bNiceMatrix}$
	};
	\node[left = -0.1cm of CXmatA, minimum height = 0.15cm] (Output){
	$\begin{bNiceMatrix}[small]1 \\ 0 \\ 0 \\ 0 \\ 0 \\ 0 \\ 0 \\ 0 \end{bNiceMatrix}^\top$
	};
	\node[right = 0.3cm of Input, minimum height = 0.15cm] (Result){
	$\frac{1}{\sqrt{2}}$
	};
	\node (EQ) at ($(CXmatA.west)!0.2!(Output.east)$) {
		$\cdot$
	};
	\node (EQ) at ($(CXmatA)!0.5!(CXmatB)$) {
		$\cdot$
	};
	\node (EQ) at ($(CXmatB)!0.4!(Hmat)$) {
		$\cdot$
	};
	\node (EQ) at ($(Hmat.east)!0.5!(Input.west)$) {
		$\cdot$
	};
	\node (EQ) at ($(Hmat.east)!0.5!(Input.west)$) {
		$\cdot$
	};
	\node (EQ) at ($(Input.west) !0.5! (Result.east)$) {
		$=$
	};
	\end{tikzpicture}}
	\caption{Matrix-vector representation}
	\label{fig:ghz_circuit_mv}
	\end{subfigure}
	\caption{Computation of the $\alpha_{000}$ amplitude of a GHZ state}
	\label{fig:ghz_circuit_and_mv}
	\vspace{-0.5cm}
\end{figure}
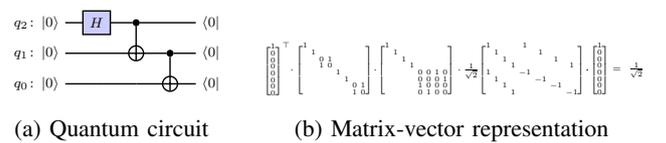

\begin{example}\label{ex:ghz}
Consider the $3$-qubit quantum circuit shown in~\autoref{fig:ghz_circuit}. 
First, a single-qubit Hadamard operation (indicated by a box labelled~$H$) is applied to the top-most qubit.
Then, two controlled-NOT gates (with the control and target qubit being indicated by $\bullet$ and $\oplus$, respectively) are applied that entangle all qubits.
Simulating this circuit with the all-zero initial state~$\ket{000}$ (shown on the left-hand side of~\autoref{fig:ghz_circuit}) leads to the well-known \emph{GHZ state} $\frac{1}{\sqrt{2}} (\ket{000} + \ket{111})$.
Thus, only determining the $\alpha_{000}$ amplitude of the resulting state (as indicated on the right-hand side of~\autoref{fig:ghz_circuit}) yields~$\frac{1}{\sqrt{2}}$.
The whole computation boils down to the multiplication of the vectors and matrices as illustrated in~\autoref{fig:ghz_circuit_mv}.
\end{example}

\section{Two perspectives on classical quantum circuit simulation}
\label{sec:two_perspectives_on_quantum_circuit_simulation}
Quantum computing is a multi-disciplinary field involving physicists, computer scientists, and many others. 
As a consequence, multiple different perspectives on quantum circuit simulation exist with differing motivations, terminologies, and backgrounds in mind.
In an effort to bridge the gap between communities, the following shall serve as an introduction to two
solutions aiming to tackle the inherent complexity of quantum mechanical systems.

\subsection{Tensor Networks}\label{sec:tensor_networks}
Quantum computing rests on the physical theory of quantum mechanics, which describes and accounts for the way nature behaves at the scale of atoms and sub-atomic particles. Quantum many-body physics, e.g., studies the collective behavior of interacting particles. Typically, particles close to another interact strongly, whereas particles at a distance hardly interact---inducing a notion of (topological) locality. This local structure naturally motivates modelling quantum many-body problems as \emph{tensor networks}~\cite{fannesFinitelyCorrelatedStates1992, biamonteTensorNetworksNutshell2017, bridgemanHandwavingInterpretiveDance2017}.

For our purposes, a \emph{tensor} can be understood as a \mbox{multi-dimensional} array of complex numbers. 
Here, the \emph{rank} of a tensor is its number of dimensions (or indices)
while a tensor's \emph{shape} specifies the number of elements in each dimension.
Two tensors sharing common indices can be \emph{contracted} into a single tensor by summing over repeated indices.

\begin{example}\label{ex:matmul}
	Let $A, B, C$ be matrices in $\mathbb{C}^{N\times N}$. Further, let the matrix product $C=AB$ be given by
	$
	C_{i,j} = \sum_{k=0}^{N-1} A_{i,k}B_{k,j},
	$
	with $i,j=0,\dots,N-1$.
	Then, this corresponds to the contraction of the rank-$2$ tensors $A = [A_{i,k}]$ and $B = [B_{k,j}]$ over the shared index $k$.
\end{example}

A \emph{tensor network} is a countable set of tensors connected by shared indices.
This is conveniently represented using a graphical notation, where individual tensors are represented as vertices of an undirected graph.
Edges between nodes describe shared indices of the respective tensors.

\begin{example}\label{ex:matmultensor}
	Consider again the situation as in \autoref{ex:matmul}. Then, the matrix multiplication $C = AB$ can be graphically represented as:
	\begin{center}
		\begin{tikzpicture}
			\node[draw] (C) {$C$};
			\node[draw, right = 2 of C] (A) {$A$};			
			\node[draw, right = 1 of A] (B) {$B$};
			\draw[] (C) -- ++ (-0.5, 0) node[left] {$i$};
			\draw[] (C) -- ++ (0.5, 0) node[right] {$j$};
			\draw[] (A) -- ++ (-0.5, 0) node[left] {$i$};
			\draw[] (B) -- ++ (0.5, 0) node[right] {$j$};
			\draw[] (A) -- (B) node[above, midway] {$k$};
			\node[] at ($(C)!0.5!(A)$) {$=$};			
		\end{tikzpicture}
	\end{center}
\end{example}

Extracting useful information from a tensor network typically requires pairwise contraction of all its individual tensors until only a single tensor remains.
In this regard, the goal is to choose an order of contractions that keeps the dimensions of contracted indices (also referred to as \emph{bond dimension}) and the shape of intermediate tensors as small as possible, because the computational effort required to contract two tensors directly correlates with these quantities. Whenever the bond dimension and the size of intermediate tensors can be kept moderate, simulation can be conducted efficiently using tensor networks. The order in which the tensors of a given network are contracted is called \emph{contraction plan}.

Based on this, there is an intuitive translation between a quantum circuit and a tensor network. 
To this end, the initial state vector and the output state vector (if only individual amplitudes should be determined) are typically represented by sets of \mbox{rank-$1$} tensors.
In addition, each $k$-qubit gate is represented by a \mbox{rank-$2k$} tensor\footnote{This is just a particular way of shaping the underlying $2^k\times 2^k$-dimensional gate matrix. Such a tensor can be understood as having $k$ ingoing and $k$ outgoing edges, each having dimension $2$.}, that is connected to the tensors preceding/following it via shared indices.
Simulating the quantum circuit then corresponds to contracting the resulting tensor network according to some contraction plan.

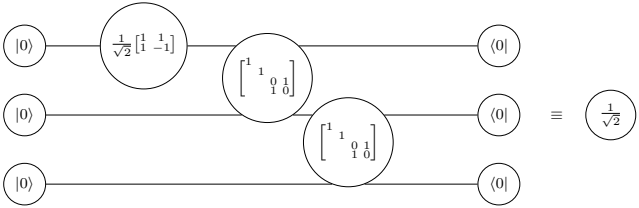
\begin{figure}
\begin{center}
\resizebox{0.95\linewidth}{!}{
		\begin{tikzpicture}
			\node[circle, draw] (q2) {$\ket{0}$};
			\node[circle, draw, below = 0.5 of q2] (q1) {$\ket{0}$};
			\node[circle, draw, below = 0.5 of q1] (q0) {$\ket{0}$};
			\node[circle, draw, right = 1 of q2] (H) {$\frac{1}{\sqrt{2}}\begin{bNiceMatrix}[small]1 & 1\\ 1 & -1\end{bNiceMatrix}$};
			\node (CXA) at (4.5,-0.6) [circle, draw]  {$\begin{bNiceMatrix}[small]1 & & &\\ & 1 & & \\ & & 0 & 1\\ & & 1 & 0\end{bNiceMatrix}$};
			\node (CXB) at (6,-1.8) [circle, draw]  {$\begin{bNiceMatrix}[small]1 & & &\\ & 1 & & \\ & & 0 & 1\\ & & 1 & 0\end{bNiceMatrix}$};	
			\node [circle, draw, right = 8 of q2] (q2o)  {$\bra{0}$};
			\node [circle, draw, right = 8 of q1] (q1o)  {$\bra{0}$};
			\node [circle, draw, right = 8 of q0] (q0o)  {$\bra{0}$};
			\node [circle, draw, right = 10 of q1] (result) {$\frac{1}{\sqrt{2}}$};
			\draw (q2) -- (H);
			\coordinate(cpa) at (intersection 1 of CXA and q2o--H);
			\draw (H) -- (cpa);
			\coordinate(cpb) at (intersection 1 of CXA and H--q2o);
			\draw (cpb) -- (q2o);
			
			\coordinate(cpc) at (intersection 1 of CXA and q1o--q1);
			\draw (q1) -- (cpc);			
			\coordinate(cpd) at (intersection 1 of CXB and q1--q1o);
			\draw (cpd) -- (q1o);	
			\coordinate(cpe) at (intersection 1 of CXA and q1--q1o);
			\coordinate(cpf) at (intersection 1 of CXB and q1o--q1);
			\draw (cpe) -- (cpf);

			\coordinate(cpx) at (intersection 1 of CXB and q0o--q0);
			\draw (q0) -- (cpx);
			\coordinate(cpz) at (intersection 1 of CXB and q0--q0o);
			\draw (cpz) -- (q0o);
			\node  (EQ) at ($(q1o.west) !0.5!(result.east)$) {
				$\equiv$
			};
			
		\end{tikzpicture}}
	\end{center}
	\caption{Tensor network for the computation from~\autoref{fig:ghz_circuit}}
	\label{fig:ghz_circuit_tn}
\end{figure}

\begin{example}\label{ex:circuit_tn}
Consider again the computation shown in~\autoref{fig:ghz_circuit}.
\autoref{fig:ghz_circuit_tn} shows how this translates to a tensor network.
To this end, each individual tensor is illustrated by a ``bubble'' containing the actual tensor data.
Note, that this representation only requires a linear amount of memory with regard to the total number of qubits and gates (in contrast to the exponential representation shown in~\autoref{fig:ghz_circuit_mv}).
\end{example}

\subsection{Decision Diagrams}
\label{sec:decision_diagrams}

In general, the representations of quantum states and operations are exponentially large with respect to the number of qubits. \mbox{Straight-forwardly} representing these entities quickly becomes limited by the sheer amount of memory required to even store them---even when resorting to supercomputing clusters.
However, the respective vectors and matrices frequently are sparse or have inherent redundancies in their representation. 
\emph{Decision diagrams}~\cite{niemannQMDDsEfficientQuantum2016, chin-yungExtendedXQDDRepresentation2011, zulehnerHowEfficientlyHandle2019, viamontesHighperformanceQuIDDBasedSimulation2004}\footnote{In the following, we mainly focus on decision diagrams as proposed in~\cite{zulehnerHowEfficientlyHandle2019}. However, the obtained findings can easily be extrapolated to other types of decision diagrams.} have been proposed as a means to exploit these redundancies in the underlying representation, which allows them to compactly represent and efficiently manipulate quantum states and operations in many cases.

For our purposes, a decision diagram is a directed, acyclic graph with complex edge weights.
A given state vector can be recursively decomposed into sub-vectors according to
\begin{gather*}
        [\alpha_{0\ldots 0}, \hdots, \alpha_{1\ldots 1}]^\top \\
        [\alpha_{0\dots}]^\top \qquad\qquad\qquad [\alpha_{1\dots}]^\top \\
        [\alpha_{00\dots}]^\top \quad [\alpha_{01\dots}]^\top \qquad [\alpha_{10\dots}]^\top \quad [\alpha_{11\dots}]^\top 
\end{gather*}
until only individual amplitudes remain. 
The resulting structure has $n$ levels of nodes, labelled $n-1$ down to $0$.
In a path from the top to the bottom, each node at level $i$ represents a \emph{decision} with two choices---whether the path leads to an amplitude where qubit $i$ is in state $\ket{0}$ or $\ket{1}$.  
Sub-vectors only differing by a constant factor can be represented by the same node in the decision diagram.
In general, this is handled by employing normalization schemes that guarantee canonicity and using hash tables to track unique nodes.

This decomposition scheme can naturally be extended to the representation of quantum gates, by adding a second dimension.
This corresponds to recursively splitting the respective matrices into four equally-sized sub-matrices according to the linear operator basis
\[
\left\{\begin{bNiceMatrix}[small]1&0\\0&0\end{bNiceMatrix}, \begin{bNiceMatrix}[small]0&1\\0&0\end{bNiceMatrix}, \begin{bNiceMatrix}[small]0&0\\1&0\end{bNiceMatrix}, \begin{bNiceMatrix}[small]0&0\\0&1\end{bNiceMatrix}\right\} = \{\ket{0}\bra{0}, \ket{0}\bra{1}, \ket{1}\bra{0}, \ket{1}\bra{1}\}.
\]
In this fashion, rather compact representations for many quantum states and operations can be obtained.

Furthermore, most operations on vectors and matrices---multi\-plication, addition, inner/outer product, tensor product, etc.---can naturally be translated to decision diagrams due to their recursive nature.
However, instead of scaling with the vectors' and matrices' dimensions, operations on decision diagrams scale with the number of nodes of the respective decision diagrams.
Hence, as long as the involved decision diagrams remain compact, computations can be conducted very efficiently. 

\begin{example}\label{ex:ddmult}
As described above, modifying the state of a quantum system by applying an operation entails the \mbox{matrix-vector} multiplication of the operation's matrix with the current state vector.
This operation can be recursively broken down according to
\begin{align*}
\begin{bmatrix}
U_{00} & U_{01} \\
U_{10} & U_{11} \\
\end{bmatrix}
\times
\begin{bmatrix}
\alpha_{0\ldots} \\
\alpha_{1\ldots} \\
\end{bmatrix}
=
\begin{bmatrix}
(U_{00} \cdot \alpha_{0\ldots}) + (U_{01} \cdot \alpha_{1\ldots}) \\
(U_{10} \cdot \alpha_{0\ldots}) + (U_{11} \cdot \alpha_{1\ldots}) \\
\end{bmatrix},
\end{align*}
with $U_{ij}\in\mathbb{C}^{2^{n-1}\times 2^{n-1}}$ and $\alpha_{i\ldots}\in\mathbb{C}^{2^{n-1}}$ for $i,j\in\{0,1\}$.
In the respective decision diagrams, 
$U_{ij}$ and $\alpha_{i\ldots}$ directly correspond to the successors of a matrix and a vector node.
As a consequence, the complexity of the multiplication scales with the product of the respective number of nodes.
\end{example}

Similarly to tensor networks, simulating a quantum circuit using decision diagrams first entails constructing decision diagram representations of the initial state vector and the individual gate matrices.
Then, multiplying these decision diagrams accordingly yields a decision diagram representation of the resulting state vector.
We refer to the order in which the individual multiplications are performed as a \emph{simulation path}.
Individual amplitudes can be efficiently extracted from the final decision diagram by accumulating the edge weights along a single path from the root node to its terminal.

\begin{figure}[t]
	\centering
	\resizebox{0.99\linewidth}{!}{
	\begin{tikzpicture}
	
	\node[label={[xshift=0.5cm, yshift=-0.5cm]}, rectangle, draw, minimum height = 3.65cm] (G) {
		\usebox{\ddfullstate}
	};
	
	\node[left = 0.01cm of G] (EQ) {
		$=$
	};
	
	\node[left = 0.05cm of EQ,label={[xshift=0.35cm, yshift=-0.5cm]}, rectangle, draw, minimum height = 3.65cm] (Initial State) {
		\usebox{\ddinitialstate}
	};	
	
	\node[left = 1.5cm of EQ,label={[xshift=0.35cm, yshift=-0.5cm]}, rectangle, draw, minimum height = 3.65cm] (H) {
		\usebox{\Hgate}
	};
	
	\node[left = 2.8cm of EQ,label={[xshift=0.35cm, yshift=-0.5cm]}, rectangle, draw, minimum height = 3.65cm] (CX) {
		\usebox{\CXgatea}
	};
	
	\node[left = 4.5cm of EQ,label={[xshift=0.35cm, yshift=-0.5cm]}, rectangle, draw, minimum height = 3.65cm] (CX) {
		\usebox{\CXgateb}
	};
	
	\node[right = 0.5cm of G,label={[xshift=0.35cm, yshift=-0.5cm]}, rectangle, draw, minimum height = 3.65cm] (Calc){$\begin{matrix} \textcolor{blue}{1} \\ \textcolor{blue}{\times} \\ \textcolor{blue}{\frac{1}{\sqrt{2}}} \\ \textcolor{blue}{\times} \\ \textcolor{blue}{1} \\ \textcolor{blue}{\times} \\ \textcolor{blue}{1} \\ \textcolor{blue}{=} \\ \textcolor{blue}{\alpha_{000}} \end{matrix}$};
			
	\end{tikzpicture}}
	\caption{DDs for the computation from~\autoref{fig:ghz_circuit_mv}}
	\label{fig:ghz_circuit_dd}
\end{figure}
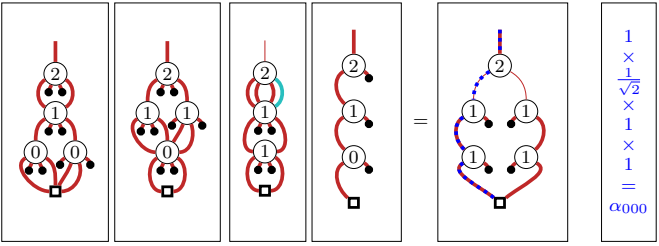

\begin{example}\label{ex:ghz_dd}
Consider again the computation shown in~\autoref{fig:ghz_circuit_mv}. \autoref{fig:ghz_circuit_dd} shows how this translates to decision diagrams.
To this end, the visualization proposed in~\cite{willeVisualizingDecisionDiagrams2021} is used, where thickness and color of an edge represent the edge weight's magnitude and phase, respectively.
Observe, that all representations have a linear number of nodes with respect to the number of qubits in contrast to the exponential representation shown in~\autoref{fig:ghz_circuit_mv}.
\end{example}

\section{Systematic Analysis}
\label{sec:two_sides_of_the_same_coin?}
Having examined these two complementary data structures, the \mbox{objective} in the remainder of this work is to establish guidelines for choosing an appropriate data structure for classical quantum circuit simulation depending on different use cases.
In the following, we shed light on this matter by systemically analyzing both techniques with regard to their most applicable \emph{abstraction level}, the \emph{desired simulation output}, the impact of the \emph{computation order}, and the ease of \emph{distributing the workload}.

\subsection*{Abstraction Level}

Several use cases for classically simulating quantum circuits on different abstraction layers exist---encompassing the development or testing of (high-level) applications and the (low-level) verification of quantum computers. Each of these abstraction layers comes with its own characteristics, e.g., various granularities of gate sets or considerations of specific hardware limitations. 

Motivated by the limited connectivity of many currently available quantum processors, near-term quantum algorithms are generally designed on lower abstraction levels, i.e., circuits typically only contain single- and two-qubit gates and feature rather local interactions.
Tensor networks are perfectly suitable in this case due to them capturing the \emph{topological} structure of a quantum circuit. 
Any such operation only acting on a couple of qubits can be represented by a compact tensor.
In contrast, the current state of the art in decision diagram-based simulation requires every decision diagram representing an operation to be extended to the full system size by forming appropriate tensor products with decision diagrams representing the identity. 
While this yields a dependency on the number of qubits, the resulting decision diagrams are still compact in general---commonly requiring only a linear number of nodes.

\begin{figure}[t]
	\centering
	\begin{subfigure}[b]{0.45\linewidth}
	\centering
	\resizebox{0.45\linewidth}{!}{
	\begin{tikzpicture}
	
	\node[label={[xshift=0.5cm, yshift=-0.5cm]}] (G) {
		\usebox{\CXgatemqubits}
	};
			
	\end{tikzpicture}}
	\caption{CX$(q_{n-1}, q_0)$}
 	\label{fig:cnotdd}
	\end{subfigure}%
	\begin{subfigure}[b]{0.55\linewidth}
	\centering
	\resizebox{0.3\linewidth}{!}{
	\begin{tikzpicture}
	
	\node[label={[xshift=0.5cm, yshift=-0.5cm]}] (G) {
		\usebox{\CXgatemulti}
	};
			
	\end{tikzpicture}}
	\caption{MCX$(q_{n-1}, \dots, q_1, q_0)$}
	\label{fig:mcxdd}
	\end{subfigure}
	\caption{DDs for common operations}
	\label{fig:dds_multiqubitoperations}
\end{figure}
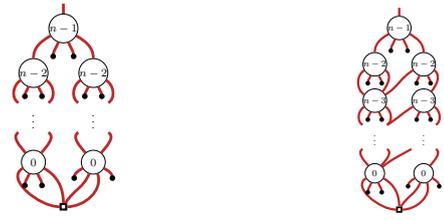

\begin{example}\label{ex:lowlevel}
	Consider an $n$-qubit system and assume that a controlled-NOT operation shall be applied with $q_{n-1}$ and $q_0$ acting as control and target, respectively. The corresponding tensor just consists of the well-known $4\times 4$ matrix representation of the controlled-NOT operation, i.e., 
	\begin{center}
		\begin{tikzpicture}
			\node[draw] (C) {$\begin{smallmatrix}
				1 & 0 & 0 & 0 \\ 0 & 1 & 0 & 0 \\ 0 & 0 & 0 & 1 \\ 0 & 0 & 1 & 0
			\end{smallmatrix}$};
			\draw[] (C.0) ++ (0, 0.3) -- ++ (1., 0) node[right] {$q_{n-1}^{\mathit{out}}$};
			\draw[] (C.0) ++ (0, -0.3) -- ++ (1., 0) node[right] {$q_{0}^{\mathit{out}}$};
			\draw[] (C.180) ++ (0, 0.3) -- ++ (-1., 0) node[left] {$q_{n-1}^{\mathit{in}}$};
			\draw[] (C.180) ++ (0, -0.3) -- ++ (-1., 0) node[left] {$q_{0}^{\mathit{in}}$};
		\end{tikzpicture}
	\end{center}
	On the other hand, the extension to the full system size has the form 
	\[P_0\otimes \mathbb{I}^{\otimes (n-2)} \otimes \mathbb{I} + P_1 \otimes \mathbb{I}^{\otimes (n-2)} \otimes X\]
	with $P_i = \ketbra{i}{i}$ denoting the projections on $0$ and $1$, respectively. 
	The corresponding decision diagram representation consists of $2n-1$ nodes, as illustrated in \autoref{fig:cnotdd}.
\end{example}

Similar to the way classical programming languages allow to abstract from assembly or machine code, higher levels of abstractions are used for developing quantum algorithms.
Descriptions of such algorithms comprise building blocks such as (Boolean) oracles, basis transformations, or problem encodings. 
While the functionality of such building blocks typically possesses some kind of structure, this structure is not necessarily topological or local.
Since any operation acting on $k$ qubits is inherently represented by a $2^k\times 2^k$ matrix, methods based on tensor networks cannot represent these constructs efficiently for large $k$.
In contrast, decision diagrams aim to exploit \emph{structural redundancies} in the underlying representations which frequently allows them to efficiently represent high-level, non-local circuit elements.

\begin{example}\label{ex:toffoli}
	Consider the simulation of the famous Grover algorithm, which provides a quadratic speed-up over classical techniques for unstructured search problems~\cite{groverFastQuantumMechanical1996}. 
	An essential part of Grover's algorithm is an \emph{oracle} that encodes solutions to the considered problem.
	Such an oracle \enquote{marks} desired states and is described by a unitary matrix $U_f$ that acts in the following way:
	\[
	U_f \ket{x}\otimes\ket{0} = 
	\begin{cases}
		\ket{x}\otimes\ket{1} & \text{if $x$ is the state searched for,} \\
		\ket{x}\otimes\ket{0} & \text{otherwise.} 
	\end{cases}
	\]
	One of the simplest cases is the search for the all-one state $\ket{1\dots 1}$.
	An oracle for this task is given by a \mbox{multi-controlled} NOT (or \emph{Toffoli}) gate which flips the value of a target qubit only if all of its control qubits are in state $\ket{1}$.
	The corresponding unitary matrix $U_f$ is given by \mbox{$\mathbb{I}\otimes \cdots \otimes\mathbb{I} \otimes X$}. 
	As a consequence, the memory required to store this matrix as a tensor grows exponentially with the system size.
	In contrast, a linear number of nodes suffices to represent it as a decision diagram, as illustrated in \autoref{fig:mcxdd}.
\end{example}

\textbf{Conclusion:} Tensor networks fit well for low-level applications of classical circuit simulation due to them capitalizing on inherent \emph{topological structure}. In contrast, decision diagrams exploit \emph{structural redundancies} in the underlying representation, which is independent of the abstraction level. As a result, decision diagrams show clear benefits for high-level use cases while, at the same time, covering low-level applications.

\subsection*{Desired Simulation Output}

The desired output of a classical quantum circuit simulation can be anything from a scalar quantity (e.g., an individual amplitude) to the complete state vector. 
In the following, the implications of the desired quantity on the performance of classical simulation methods based on tensor networks and decision diagrams are studied.

First, assume that a representation of the complete output state vector of a quantum circuit shall be computed.
Using tensor networks, this amounts to contracting the whole tensor network into a single rank-$n$ tensor of size~$2^n$.

Inevitably, this incurs the same exponential memory requirement as for the straight-forward \mbox{matrix-vector} calculation shown in \autoref{sec:quantum_circuit_simulation}.
As a consequence, it is infeasible in general to compute the complete output state vector (or, more generally, exponentially many amplitudes) using tensor networks. 
Many specialized types of tensor networks, such as \emph{Matrix Product States} (MPS), \emph{Tree Tensor Networks} (TTN), \emph{Multi-scale Entanglement Renormalization Ansatz} (MERA), or \emph{Projected Entangled Pair States} (PEPS), have been proposed that try to impose some structure in the whole state representation by decomposing it into smaller tensors (see~\cite{viamontesCheckingEquivalenceQuantum2007,ciracMatrixProductStates2021} and the references therein).
However, these techniques typically aim for a slightly different goal than considered in this work---namely, efficiently \emph{approximating} the state of quantum mechanical systems.

Using decision diagrams, individual quantum operations can commonly be represented in linear size (as discussed in the previous section).
Computing a representation of the full output state then entails the subsequent multiplication of the respective decision diagrams.
It has been shown, that, for many practically relevant cases, the size of the (intermediate) diagrams during the simulation remains polynomial with respect to the number of qubits~\cite{zulehnerAdvancedSimulationQuantum2019}.
Consequently, the inherent exponential complexity is alleviated in those cases.
Once the final decision diagram is obtained, an arbitrary number of amplitudes can be extracted with little overhead by recursively traversing the decision diagram and accumulating edge weights along the way.

On the other end of the spectrum, it might be desirable to determine a single scalar quantity, e.g., an individual amplitude or the expected value of some observable.
For methods based on tensor networks, this is accomplished by fixing the output indices of the circuit's tensor network.
In contrast to the calculation of the complete output state, contracting the resulting tensor network results in a single rank-0 tensor, i.e., a scalar (as, e.g., in \autoref{fig:ghz_circuit_tn}).
Whenever a contraction plan can be employed that keeps the size and bond dimension of intermediate tensors in check, individual amplitudes can be determined very efficiently.
Instead, the current state of the art in decision diagram-based simulation does not facilitate such a reduction in complexity.
The complete output state vector always needs to be determined independently of whether the entire state or only a single amplitude is desired (as illustrated in \autoref{fig:ghz_circuit_dd}).\medskip

\textbf{Conclusion:} 
While a general conclusion cannot be drawn without prior assumptions, the following tendency stands out:
Tensor networks should be preferred for computing scalar quantities as the result of a quantum circuit simulation---given that a suitable contraction plan can be determined.
Decision diagrams are more suited towards full state vector simulation---provided that the involved decision diagrams remain compact throughout the simulation.

\subsection*{Determining the Order of Computation}

\begin{table*}[t]
\centering
\caption{Summary of main results and resulting guidelines}\vspace*{-1mm}
\label{tab:sumtable}
\resizebox{0.95\linewidth}{!}{
\begin{tabular}{ lll }
  & \textbf{Tensor Networks} & \textbf{Decision Diagrams}\\
 \midrule
 \textbf{Abstraction level}  & Efficient for low-level applications & Independently applicable \\ & through exploiting topological structure & through exploiting structural redundancies \\ &  & with clear benefits for high-level applications \medskip\\
 \textbf{Desired simulation output} & Should be preferred for computing scalar quantities & Should be preferred for computing the full state vector \\ & given a suitable contraction plan can be determined & given that the involved decision diagrams remain compact \medskip\\ 
 \textbf{Computation order} 
 & A plethora of methods is available aiming to & First methods have been developed aiming to \\ &  keep the size and shape of the tensors in check & keep intermediate decision diagrams as compact as possible\medskip\\
 \textbf{Distributing the workload}  
 & Vast options are available for simulations & Potential has been demonstrated, \\ & near the supremacy threshold on HPC systems & but solutions beyond desktop computers are yet to be explored
 \\\bottomrule 
\end{tabular}
}
\end{table*}

Given a particular quantum circuit simulation, there exists an immediate duality between the contraction plan for tensor networks, i.e., the order in which the tensors are contracted, and the simulation path for decision diagrams, i.e., the sequence in which the individual multiplications are performed.
Both techniques have in common that their performance heavily depends on the order in which the individual operations are conducted.
This problem has been intensively studied for tensor networks~\cite{chi-chungOptimizingClassMultidimensional1997, grayHyperoptimizedTensorNetwork2021, huangClassicalSimulationQuantum2020, boixoSimulationLowdepthQuantum2018, lykovTensorNetworkQuantum2020}.
While not as extensive as for tensor networks, research towards this direction has also been conducted for decision diagrams~\cite{zulehnerMatrixVectorVsMatrixMatrix2019, burgholzerAdvancedEquivalenceChecking2021, burgholzerVerifyingResultsIBM2020, burgholzerExploitingArbitraryPaths2022}---in particular, for the use case of equivalence checking.
Thus, in the following, we consider equivalence checking as the perfect use case for answering the question, whether the strategies for determining the order of computation are interchangeable between domains.

To this end, given two quantum circuits $G$ and $G'$, it shall be decided, whether simulating both circuits yields equivalent outputs for a fixed input~$\ket{\varphi}$. 
One possible solution is to simulate both circuits with input $\ket{\varphi}$ and compare the resulting states.
However, in general, this is neither feasible for tensor networks (due to the inherent exponential memory requirement independent of the contraction plan), nor for decision diagrams (due to their worst case exponential complexity independent of the simulation path).

Due to the reversible nature of quantum computing, there is a more promising approach for checking the equivalence of both circuits.
To this end, one of the circuits is inverted and concatenated with the other circuit---forming a new circuit $\tilde{G}=G\,G^{\prime -1}$.
Whenever both original circuits are indeed equivalent for input~$\ket{\varphi}$, simulating the circuit $\tilde{G}$ with input $\ket{\varphi}$ does not affect the state at all, i.e., it holds that $|\expval{\tilde{G}}{\varphi}|^2 = 1$.
Now, the result of the computation is not a (potentially) exponential representation, but a single complex number.
As a consequence, the order of operations to arrive at this compact result becomes crucial.

Due to the topological structure of tensor networks, their performance is independent of the actual content of the individual tensors, but only depending on their size and shape.
As a result, precise estimates for the required amount of memory and floating point operations for contracting a tensor network according to a particular contraction plan can be derived a-priori, i.e., without conducting the actual contraction.
This makes it possible to explore the vast search space of contraction plans with the goal of keeping the dimension of shared indices and the sizes of the intermediate tensors during the contraction as small as possible.
However, determining an optimal contraction plan is an extremely delicate process, and has even been shown to be NP-hard~\cite{chi-chungOptimizingClassMultidimensional1997}.
Accordingly, a plethora of heuristic methods have been proposed to find efficient contraction plans for tensor networks~\cite{grayHyperoptimizedTensorNetwork2021, huangClassicalSimulationQuantum2020, boixoSimulationLowdepthQuantum2018, lykovTensorNetworkQuantum2020}.
In general, these techniques start from the \enquote{outside}, i.e., the small tensors of the initial state and the output amplitudes, and \enquote{work their way inwards} with the goal of keeping the size of intermediate tensors in check.
While estimating the complexity of the multiplication of two decision diagrams is straight-forward (it scales with the product of the decision diagrams' nodes), estimating the size of the resulting decision diagram is extremely difficult in the general case without performing the actual multiplication (due to decision diagrams heavily relying on the amount of redundancy being present in the resulting representation). 
As a result, no clear runtime estimation can be derived for classical simulation based on decision diagrams.
However, it has been shown in~\cite{burgholzerExploitingArbitraryPaths2022}, that translating tensor network contraction plans to simulation paths for decision diagrams can allow for speedups of up to several orders of magnitude compared to the established simulation approach.

The efficiency of a simulation path for decision diagrams heavily depends on the amount of redundancy present in the underlying representations of intermediate results.
While not as extensive as for tensor networks, it has been shown that choosing the right simulation path can make the difference between linear and exponential runtime as well as space~\cite{zulehnerMatrixVectorVsMatrixMatrix2019, burgholzerAdvancedEquivalenceChecking2021, burgholzerVerifyingResultsIBM2020, burgholzerExploitingArbitraryPaths2022}.
The resulting strategies mainly achieve their efficiency from incorporating certain characteristics about the considered simulation task.
For the problem at hand, the quintessence is to try to keep the decision diagrams occurring during the computation as close as possible to the identity\footnote{The identity constitutes the best case for a matrix decision diagram as it only requires a single node per qubit and has no non-trivial edge weights.}.
This is accomplished by starting \enquote{in between} $G$ and $G^{\prime -1}$ and \emph{alternating} between applications of gates from $G$ and (inverted) gates from $G'$ according to some scheme.
Whenever a scheme,~i.e., a simulation path, can be employed that allows to keep the intermediate decision diagrams close to the identity, the equivalence of both circuits can be concluded efficiently. 

While the same strategy could be employed to contract the corresponding tensor network, this will most certainly not result in an efficient contraction plan.
This is due to the fact that, sooner or later, all qubits will be involved in the computation---leading to a maximally-large tensor of size $2^n\times 2^n$.
Therefore, the opposite direction, i.e., translating a simulation path to a corresponding contraction plan, can hardly be expected to be effective.\medskip

\textbf{Conclusion:} 
Contraction plan and simulation path seem very similar from an outside perspective. But, when taking a closer look, it is clear that they are trying to achieve different goals. 
An efficient contraction plan tries to keep the size and shape of intermediate tensors in check. In contrast, the goal of simulation paths is to have intermediate decision diagrams which are as compact as possible. 
While translating an efficient contraction plan for tensor networks into a simulation path has been demonstrated to yield speedups of up to several orders of magnitude, the inverse, i.e., translating an efficient simulation path for decision diagrams into a contraction plan, can hardly be expected to be effective.

\subsection*{Distributing the Workload}
In order to perform classical simulations of large quantum circuits, the respective methods have to be designed to run on high performance supercomputing clusters.
Hence, it is key to distribute the workload of a particular computation to multiple (heterogeneous) \enquote{workers} in order to fully utilize the available resources.

To this end, the task of contracting two tensors is inherently parallelizable on CPUs as well as GPUs due to its regular structure.
On the other hand, it has been shown in~\cite{hillmichConcurrencyDDbasedQuantum2020} that parallelizing individual decision diagram operations is not straight-forward (not even on CPUs).
Intuitively, the fashion in which decision diagrams try to exploit redundancies mitigates the potential of trivial parallelization as it exists for regular matrices and vectors.

Given a specific contraction plan, independent contractions can be efficiently conducted in parallel~\cite{vincentJetFastQuantum2021}.
Due to the immediate duality between contraction plans and simulation paths witnessed in the previous section, this also holds for decision diagrams.
However, some additional effort is required to consistently maintain multiple decision diagram packages in parallel.

Furthermore, methods based on tensor networks employ \emph{slicing} or \emph{cutting} techniques, which essentially subdivide a tensor network by fixing the values of certain indices~\cite{grayHyperoptimizedTensorNetwork2021, villalongaFlexibleHighperformanceSimulator2019, chen64qubitQuantumCircuit2018}. 
The overall contraction then reduces to the sum of several independent contractions of simpler tensor networks.
Promising initial investigations towards realizing such schemes for decision diagrams have been conducted in~\cite{burgholzerHybridSchrodingerFeynmanSimulation2021}.\medskip

\textbf{Conclusion:}
Significant efforts have been conducted to scale methods based on tensor networks to the realm of exascale computing~\cite{nguyenTensorNetworkQuantum2021, brennanTensorNetworkCircuit2021, vincentJetFastQuantum2021}.
In contrast, corresponding solutions for decision diagrams are still in their infancy and the current state of the art is rather tailored for desktop systems---its limits yet to be explored.
As a consequence, classical computations near the quantum supremacy threshold will probably be tackled using tensor networks for the foreseeable future, while decision diagrams are a suitable means to classically simulate quantum circuits on desktop computers, e.g., during application development. 

\section{Conclusion}\label{sec:conclusion}
In this paper, we reviewed and discussed two complementary approaches to tackle the inherent complexity of classically simulating quantum circuits---tensor networks and decision diagrams.
To this end, we systematically analyzed
both techniques with regard to their most applicable \emph{abstraction level}, the \emph{desired simulation output}, the impact of the \emph{computation order}, and the ease of \emph{distributing the workload}.
\autoref{tab:sumtable} summarizes the main results of this analysis, which provide guidelines that shall aid designers in deciding what data structure to choose for which use cases.

\section*{Acknowledgements}
This work received funding from the European Research Council (ERC) under the European Union’s Horizon 2020 research and innovation program (grant agreement No. 101001318), was part of the Munich Quantum Valley, which is supported by the Bavarian state government with funds from the Hightech Agenda Bayern Plus, and has been supported by the BMWK on the basis of a decision by the German Bundestag through project QuaST, as well as by the BMK, BMDW, and the State of Upper Austria in the frame of the COMET program (managed by the FFG).

\printbibliography

\end{document}